\pdfoutput=1
\documentclass[aps,pra,showpacs,reprint,groupedaddress]{revtex4-1}
\usepackage{amsmath}
\usepackage{epsfig}
\usepackage{epstopdf}
\begin{document}
\title[Universal single-server blind quantum computation for classical clients]{Universal single-server blind quantum computation for classical clients
}
\author{Hai-Ru Xu$^1$}%

\author{Bang-Hai Wang$^1$}%
\email{bhwang@gdut.edu.cn}
\affiliation{%
$^1$School of Computers, Guangdong University of Technology, Guangzhou 510006, People's Republic of China
}%
\date{\today}

\begin{abstract}
Blind quantum computation allows a client without enough quantum technologies
to delegate her quantum computation to a remote quantum server, while keeping her input, output and algorithm secure. In this paper, we propose a universal single-server and almost-classical-client blind quantum computation protocol. In this protocol, the client interfaces with only one server and the only ability of the client required is to get particles from the trusted center and forward them to the server. We present an analysis of this protocol and modify it to a universal single-server and fully-classical-client blind quantum computation protocol by improving the ability of the trusted center. Based on our protocols and recent works, a new ``Cloud + Certificate Authority (CA)" style is proposed for the blind quantum computation.

\end{abstract}

\maketitle

\section{Introduction}

Quantum computation has been recognized as a novel computing model which is based on the theory of quantum mechanics.
Some experimental results show that quantum computation features a faster computing speed than classical computation and solves tough computation problems for classical computers~\cite{PWSHOR,LKG,Nielsen}.
Blind quantum computation (BQC), which combines notions of quantum cryptography and quantum computation, was put forward to help clients without enough quantum power to acquire quantum computation with keeping their data and algorithm private~\cite{Childs,BFK,BKBFZW,MFujii,MVFMBQC,MDF,TMKF,Qinli,Sheng}.

The idea of BQC was derived from Childs' work in 2005~\cite{Childs}. The Childs' work uses the quantum circuit model and requires the client to own quantum memory and to be able to implement the SWAP gate. Broadbent, Fitzsimons and Kashe (BFK) presented the first universal BQC protocol in 2009~\cite{BFK}. Based on measurement-based quantum computation, the BFK protocol does not require the client to have any other quantum power and memory than with single-qubit states~\cite{BFK}. After that, more and more improved and stronger protocols were proposed \cite{MFujii,MVFMBQC,MDF,TMKF,Qinli,Sheng,Hayashi2015,Morimae2015,Greganti2016,Kong2016,Fitzsumons2017,Sheng2018,Li2018}. Morimae and Fujii proposed a fault-tolerant BQC protocol which makes blind quantum computation more feasibility~\cite{MFujii}. Morimae put forward a protocol of verification for the measurement of BQC \cite{MVFMBQC}. Mantri et al. discussed the optimality of BQC protocols \cite{MDF}. Li et al. proposed a BQC protocol with identity authentication \cite{Li2018}. At the same time, the theoretical work are expected to be applicable to physical implementation. Recently, the BFK protocol has been demonstrated experimentally in a quantum optics setting system~\cite{BKBFZW}. We refer the reader to Ref. \cite{fitzsimons2017private} for a recent review on BQC.

It is crucial for the implementation of BQC to make the client classical. 
The BFK protocol \cite{BFK} can be upgraded to two-server BQC protocol~\cite{TMKF} with the client being classical, if the trusted center is introduced. However, communication is not allowed between the servers. To solve the servers-communication problem, Li et al. put forward a triple-server~\cite{Qinli} BQC protocol, which makes the client almost classical. That whether a fully classical client can delegate a quantum computation to an untrusted quantum server while fully maintaining privacy is one of the big open questions in quantum cryptography \cite{Dunjko2016,TMTK,Morimae2019}. For our purpose, we focus BQC with the trusted center in this paper. We first present a single-server BQC protocol makes the client almost classical without improving the ability of the trusted center. And the client can be fully classical if the trusted center can discard qubits instead of the client. Based on recent works and our protocols, we propose a new ``Cloud + CA" style for the blind quantum computation.

\section{Preliminaries}

\subsection{Measurement-based quantum computing and the technology of entanglement swapping}

\subsubsection{Measurement-based quantum computing}

Different from the quantum circuit model, where quantum computations are implemented by unitary operations, the measurement-based quantum computing is processed by employing a sequence of measurements to designated an entangled resource state, usually a cluster state which is a special graph state (a kind of multi-partite entangled state, related to a graph in mathematics) in designated bases \cite{briegel2009measurement,jozsa2006introduction,raussendorf2003measurement}. These measurements are performed on a universal resource state--the cluster state--which is independent of the algorithm to be implemented. The measurement-based quantum computing is also called one-way quantum computing because the resource state is destroyed by the measurements.




Concretely, a measurement-based quantum computation proceeds as follows (see, e.g., \cite{briegel2009measurement} and references therein). (i) A classical input is provided to specify the data and the program. (ii) A two-dimensional (2D)-cluster state $|C\rangle$, which serves as the resource for the computation, of sufficiently large size is prepared. (iii) A sequence of adaptive one-qubit measurements $M$ is performed on certain qubits in the cluster state. The measurement bases depend on the program and on the outcomes of previous measures in each step of the computation. The measurement directions, which have to be chosen in each step, is computed by a simple classical computer. (iv) The state of the system has the form $|\delta^a\rangle|\psi^a_{out}\rangle$ after the measurements, where $a$ indexes the collection of measurement outcomes of the different branches of the computation. The state $|\psi^a_{out}\rangle$ in every branch are equal to
 the desired output state up to a local operation (Pauli), and the measured qubits are in a product state $|\delta^a\rangle$, which still depends on the measurement outcomes. The measurement-based quantum is universal because any quantum computation can deterministically be realized even though the results of the measurements in each step of the computation are random.

 Besides these measurement-based BQC protocols, there are other types of BQC, for instance, BQC based on nonlocal games and self-testing of EPR pairs \cite{reichardt2013classical,huang2017experimental}.

\subsubsection{The technology of entanglement swapping}
The technology of entanglement swapping~\cite{Zukowski,Pan,Bose,Polking,Ma} is used in our protocol, as well as in the triple-sever protocol.
As one of the most significant technology of quantum mechanics, entanglement swapping is an essential resource of quantum information, which was used in quantum dense coding, quantum teleportation~\cite{Bennett,Lu}, quantum cryptography~\cite{Zhang,Xiu}, and quantum repeaters~\cite{Briegel,Duan}.
Entanglement swapping allows two or more independent systems to build up entanglement with each other by switching their photons.

Here, the four Bell states be denoted by $|\psi_{z,x}\rangle = (I\otimes X^xZ^z)(|0\rangle|0\rangle+ |1\rangle|1\rangle)/\sqrt2$, where $(z,x) \in \{0,1\}^2$, $X=|0\rangle\langle 1|+|1\rangle\langle 0|$, and $Z=|0\rangle\langle 0|-|1\rangle\langle 1|$.
Suppose there are two EPR pairs $(A,B)$ and $(A^{'}, B^{'})$ denoted in the Bell states $|\psi_{z,x}\rangle_{A,B}$ and $|\psi_{z^{'},x^{'}}\rangle_{A^{'},B^{'}}$, respectively.
If we perform joint measurement on particles $B$ ($A$) and $B^{'}$ ($A^{'}$) in the Bell basis, the particles $A$ ($B$) and $A^{'}$ ($B^{'}$) would be entangled and the combined state of them is one of the four Bell states, as shown in Fig. 1. The accurate form is determined by the result of the joint measurement outcome.

The Bell state $|\psi_{0,0}\rangle =(|00\rangle+|11\rangle)/\sqrt2$ is to perform entanglement swapping in our protocol. Two Bell states $|\psi_{0,0}\rangle_{A,B}$ and $|\psi_{0,0}\rangle_{A^{'},B^{'}}$ will be as an example to show the entanglement swapping process as follows.
The combined state of the two Bell states is $|\psi\rangle_{A,B;A^{'}, B^{'}} =|\psi_{0,0}\rangle_{a,b}\otimes|\psi_{0,0}\rangle_{A^{'},B^{'}}$. After swapping $B$ and $A^{'}$, the combined state is
\begin{eqnarray*}
|\psi\rangle_{A,B;A^{'}, B^{'}}
    &=& \frac{1}{2}(|\psi_{0,0}\rangle|\psi_{0,0}\rangle+
    |\psi_{0,1}\rangle|\psi_{0,1}\rangle +\\
    && |\psi_{1,0}\rangle|\psi_{1,0}\rangle+ |\psi_{1,1}\rangle|\psi_{1,1}\rangle)_{A,A^{'};B,B^{'}}.
\end{eqnarray*}

It is not difficult to conclude that the state of $(A,A^{'})$ ($(B,B^{'})$) will collapse to the corresponding one of the four Bell states if $(B,B^{'})$ ($(A,A^{'})$) is measured. For example, if $(B,B^{'})$ is measured in Bell basis and the result is $(0,1)$, the state of $(A,A^{'})$ will collapse to $|\psi_{0,1}\rangle$.


\begin{figure}[htbp]
\centering
\includegraphics[width=.95\columnwidth]{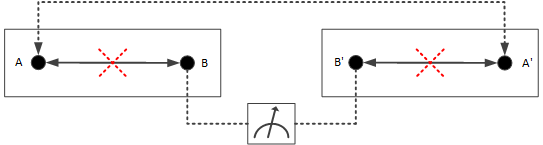}
\caption{The procedure of swapping entanglement. Two solid line segments with arrows at both ends denote connected entangled particles $(A,B)$ and $(A',B')$, respectively. The dashed line segment with arrows at both ends denotes connected entangled $(A,A')$ after the bell state measurement on $(B,
B')$ is made.} \label{fig1}
\end{figure}

\subsection{Universal BQC protocols}
Before proceeding to our protocol, we briefly review the BFK BQC protocol in Ref.~\cite{BFK}, modified double-server BQC protocol in Ref.~\cite{TMKF} and triple-server BQC protocol in Ref.~\cite{Qinli}.

Suppose that the client, Alice, requires to implement quantum computation on the $m$-qubit graph state corresponding to the graph $G$. The quantum operation that Alice wants to perform is to measure the $i$th qubit in the basis $\{|0\rangle\pm e^{i\phi_i}|1\rangle\}$, where $\phi_i\in S\equiv\{k\pi/4|k=0,1,...,7\}$. 
Let us recall the single-server BFK BQC protocol.

\textbf{Step A} Alice produces m qubits and sends them to the server, Bob. This state of single qubit is $|\theta_i\rangle=|0\rangle+e^{i\theta_i}|1\rangle (i=1,2,\cdots,m)$, where $\theta_i$ is uniformly chosen from the set $S$.

\textbf{Step B} Alice asks Bob to prepare a brickwork state based on the graph $G$ specified by her. The brickwork state is a new universal family of graph states which, unlike cluster states, only requires $(X,Y)-$plane measurements. For more details, we refer to \cite{BFK}.

\textbf{Step C} Bob generates the brickwork state $|G(\theta)\rangle$ by employing controlled-Z gates on the received qubits according to $G$.

\textbf{Step D} Alice randomly choose $r_i$ and computes $\delta_i=(\theta_i+\phi'_i+r_i\pi)$ mod $2\pi$, where $\phi'_i$ is obtained by the previous measurements and $\phi_i$, and then sent $\delta_i$ to Bob if Alice needs Bob to measure the $i$th qubit of $|G(\theta)\rangle$ for $i=1,2,\cdots,m$.

\textbf{Step E} Bob performs a measurement on the $i$th qubit in the basis $\{|\pm\delta_i\rangle\}$ for $i=1,2,\cdots,m$, and informs Alice about the measurement result.


We sketch the proof of that this protocol is unconditionally security (see Ref. \cite{BFK} for more details). The universality of the brickwork state guarantees that Bob can at most learn the dimension $(j,k)$ of the brick work state, giving an upper bound on the size of Alice's computation. Alice sends Bob classical information, $\delta_{x,y}=(\phi'_{x,y}+\theta_{x,y}+r_{x,y}\pi)$, where $\theta_{x,y}$ is chosen with a uniformly random, $r_{x,y}\in \{0,1\}$ is uniformly random, and the actual measurement angles $\phi'_{x,y}$ is a modification of the initial $\phi_{x,y}$ that depends on previous measurement outcomes. Bob gets $\delta_{x,y}$ during the protocol, but Bob does not know $r_{x,y}$ since $r_{x,y}$ is independent of everything else, and the quantum system initially sent from Alice to Bob consists of copies of the two-dimensional completely mixed state, which is fixed and independent of $\phi_{x,y}$.

The BFK BQC protocol shows that the server, Bob, cannot get any information about Alice's input, output, or algorithm. However, Alice has to prepare single qubits in the BFK (single-sever) BQC protocol.
To solve this problem, a double-server BQC protocol was presented~\cite{BFK,TMKF} in which Alice can be fully classical.
For the double-server BQC protocol~\cite{BFK,TMKF}, Alice can delegate her quantum computation to two quantum servers, Bob1 and Bob2, who share Bell pairs but cannot communicate with each other.
The trusted center prepares Bell pairs and distributes them to Bob1 and Bob2, respectively.
Alice sends classical message to Bob1. And Bob1 measures his part of Bell pair according to classical message and sends the measurement result to Alice. Alice can achieve quantum computation by running the single-server BFK protocol~\cite{BFK} with Bob2.

\begin{figure}[htbp]
\centering
\includegraphics[width=.95\columnwidth]{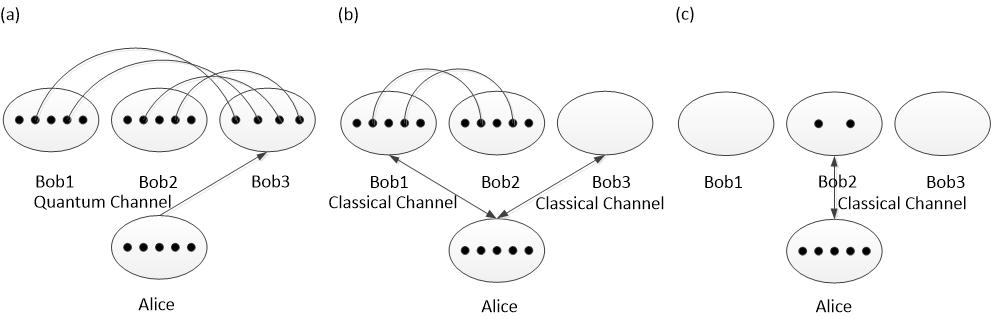}
\caption{The triple-server blind protocol. (a) Alice randomly selects and forwards qubits to Bob3. (b) Bob3 measures the qubits which are specified by Alice and returns the measurement result.
Alice sends $n$ classical message to Bob1 and
Bob1 performs measurements on his $n$ qubits and reports the outcome to Alice.
(c) Alice notifies Bob2 to hold the corresponding $m$ particles and runs the BFK protocol with Bob2.} \label{fig2}
\end{figure}

Although it is great to learn that the client could be classical in double-server protocol, the noncommunication problem between the two servers makes the protocol less practical.
In order to solve this problem, Li et al.~\cite{Qinli} proposed a triple-server BQC protocol with entanglement swapping~\cite{Zukowski,Pan,Bose,Polking,Ma}. In this protocol,
three quantum servers can communicate with each other. The client only needs to be capable of getting access to quantum channels. It can be seen that the client can delegate her quantum computation to quantum servers while keeping her data private~\cite{Qinli} (see also Fig.~\ref{fig2}).
Here, the trusted center prepares $n=2(2+\delta)m$ ($\delta$ is some fixed number) Bell pairs and distributes the first qubit of them to Bob1 and Bob2 and the second qubit to Alice. Alice randomly transmits the particles to Bob3 and records the position, or discards them.
Bob3 implements Bell state measurement on the particles according to classical information of Alice and sends the outcome to Alice.
Alice sends classical messages to Bob1 and Bob1 measures his particles in the basis according to these messages and sends the measurement results to Alice.
Based on the results, Alice sends classical information to Bob2. And Bob2 keeps a part of his qubits according to this information.
Alice can achieve quantum computation by running the single-server BFK protocol~\cite{BFK} with Bob2.


\section{Universal single-server and almost-classical-client blind quantum computation protocol and analysis}
\label{criteria}
\subsection{Universal single-server and almost-classical-client blind quantum computation protocol}
\begin{figure}[htbp]
\centering
\includegraphics[width=.95\columnwidth]{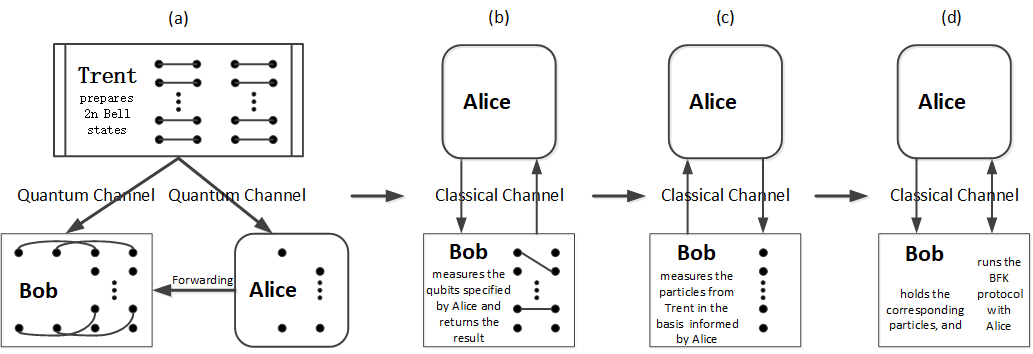}
\caption{(a) Trent prepares Bell states and distributes particles to Bob and Alice, respectively, and Alice randomly selects and forwards qubits to Bob.
(b) Bob measures the qubits which are specified by Alice and returns the result.
(c) Bob performs measurements on his first $n$ qubits from Trent in the basis informed by Alice and reports the outcome to Alice.
(d) Alice notifies Bob to hold the corresponding $m$ particles and runs the BFK protocol with Bob.} \label{fig3}
\end{figure}

Based on the previous works, we propose a single-server BQC protocol where Alice can perform her quantum computation on one quantum server Bob and keep her input, output and algorithm private.
Alice does not need any quantum power, such as owning quantum memory, performing SWAP gate, or generating any quantum state. What she need is to be capable of getting access to quantum channel.
The steps are described as follows (see also Fig.~\ref{fig3}).


\textbf{Step 1} 
The trusted center Trent prepares $2n$ Bell states, $|\psi_{0,0}(B_k,A_k)\rangle (k=1,2,\cdots,2n)$, where $n=(2+\delta)m$ ($\delta>0$ is some fixed number), and distributes the particles $B_k$ of each Bell state to Bob.
After Bob gets all the particles, Trent distributes the other particles $A_k$ to Alice.

\textbf{Step 2} Alice randomly forwards the particles $A_k$ to Bob and records the position of them, or simply discards them.
Bob places these particles in quantum registers following the incoming sequence.

\textbf{Step 3}
Suppose that the $2m$ particles are $A_{s_1},A_{s_2},\cdots,A_{s_{m}}$ and $A_{t_1},A_{t_2},\cdots,A_{t_{m}}$, where $1\leq s_i\leq n<t_i\leq 2n, i\in\{1,2,\cdots,m\}$.
Alice notifies Bob to perform Bell state measurements on particles $A_{s_i}$ and $A_{t_j}$, $(i,j)\in\{1,2,\cdots,m\}$ and gives their positions in the sequence of the particles which are from Alice. By the the technology of entanglement swapping introduced above, Bob measures the particles and submits the outcome $(z_{s_i}^{'},x_{t_j}^{'})\in \{0,1\}^2$ to Alice.

\textbf{Step 4} Based on the measurement outcome $(z_{s_i}^{'},x_{t_j}^{'})$ ($i\in\{1,2,\cdots,m\}$), Alice can know that the combined state of particles $B_{s_i}$ and $B_{t_j}$ from Trent, is just $|\psi_{z_{s_i},x_{t_j}}(B_{s_i},B_{t_j})\rangle$ and get the value of $z_{s_i}$ and $x_{t_j}$. (We refer the reader to Ref. \cite{Qinli} for more details.)

\textbf{Step 5} Alice sends $n$ classical message $\{\overset{\sim}{\theta_k}=(-1)^{x_k}\theta_k+z_k\pi\}_{k=1}^n$ to Bob, where $\theta_{k}$ is randomly selected from the set $S$. Here $(z_k,x_k)$ depends on Alice at step 4, where $k\in \{1,2,\cdots,n\}-\{s_1,s_2,\cdots,s_m\}$, $(z_k,x_k)\in \{0,1\}^2$. And $\theta_k\in S$ are selected to make $\overset{\sim}{\theta_1},\overset{\sim}{\theta_2},\cdots,\overset{\sim}{\theta_n}$ uniformly distributes over all the eight elements of the set $S$.

\textbf{Step 6} Bob measures his first $n$ qubits from Trent in the basis $\{\pm\overset{\sim}{\theta_k}\}, \{k=1,2,\cdots,n\}$, and sends the measurement outcome $\{b_k\}_{k=1}^{n}$ to Alice.
Alice simply keeps the result $(b_{s_1},b_{s_2},\cdots,b_{s_m})$ of the particles entangled with particles in another $n$ qubits from Trent.

\textbf{Step 7} Alice sends classical message $\{t_j\}_{t=1}^m$ to Bob and notifies him to keep only particles $\{B_{t_j}\}_{t=1}^m$.
Bob relabels these particles as $B_1,B_2,\cdots,B_m$ in order, and here the combined state of these particles is $\bigotimes_{i=1}^m|\theta_{s_i}+b_{s_i}\pi\rangle$.

\textbf{Step 8} Alice and Bob start the BFK protocol from the second step (\textbf{Step B} above), taking $\theta_i=\theta_i + b_{s_i}\pi, (i=1,2,\cdots,m)$.

Numbers of blind quantum computation protocols have been presented in recent years, including single-, double-, triple-, and multiple-server protocols \cite{Kong2016}. In the single-server protocol, the client needs to prepare single qubits.
Although the client can be classical in the double-server protocol, the two servers, who share Bell state from the trusted center, are not allowed to communicate with each other.
Recently, the triple-server protocol solves the noncommunication problem.
Three servers, however, make the implementation of the computation sophisticated and unrealistic. Different from the triple-server BQC protocol~\cite{Qinli}, only one server instead of three servers is involved in the proposed protocol. Comparisons among the BFK BQC protocol \cite{BFK}, the two-server BQC protocol \cite{TMKF}, the triple-server BQC protocol~\cite{Qinli}, and the proposed protocol from the number of servers, communications between servers, a trust center required or not required, and the required power of client (fully classical, quantum state preparation or quantum channel) as shown in table 1.

\begin{table*}[htbp]
\centering
\caption{Resource requirement of BQC protocols}
\begin{tabular}{c|c|c|c|c}
    \hline
    \hline
 & Number of  & Communications  &  Trust center & Quantum power of client  \\
 & servers    & between servers &  required             &   \\
    \hline
  BFK BQC protocol       & 1 & & N &Preparation of single qubits\\
    \hline
  Two-server protocol   & 2 &Prohibited &Y & No quantum ability\\
    \hline
  Triple-server protocol & 3 &Allowed & Y & Access to quantum channels  \\
      \hline
  Proposed protocol      & 1 && Y & Access to quantum channels  \\
      \hline
\end{tabular}
\label{tab1}
\end{table*}

\subsection{Analysis}
We show that this protocol is secure. Steps from Step 1 to Step 7 in this protocol can be treated as the first step in BFK protocol~\cite{BFK}. Similarly, $\theta_i + b_{s_i}\pi, (i=1,2,\cdots,m)$ is produced in these steps to replace $\theta_i$ in~\cite{BFK}.
Therefore, according to the proof of security of BFK protocol (Theorem 3)~\cite{BFK}, the protocol is secure if Bob cannot get any information about $\theta_i + b_{s_i}\pi$.
In the process, Bob can get the outcome $(z_{s_i}^{'},x_{t_j}^{'})$ and the combined state $|\psi_{z_{s_i},x_{t_j}}(B_{s_i},B_{t_j})\rangle$, but he does not know which the two corresponding particles $B_{s_i}$ and $B_{t_j}$ are.
Although Bob gets classical message $\{\overset{\sim}{\theta_k}\}_{k=1}^{n}$, which are uniformly distributed on all the eight elements of the set $S$, he still cannot get any information about $\{\theta_{s_i}\}_{i=1}^{m}$, because he has no idea about the corresponding $\{s_i\}_{i=1}^{m}$ and $\{t_j\}_{j=1}^{m}$ which are randomly selected by Alice. Then Alice keeps $\theta_i + b_{s_i}\pi$ private and makes the protocol secure.

Furthermore, if Bob sends wrong measurement result to Alice in purpose, Alice can detect the case. In Step 1, Trent generates $h$ decoy Bell states $|\psi_{z,x}(A_{h1}.A_{h2})\rangle$, where $(z,x) \in \{0,1\}^2$.
Trent sends the first particle of Bell states to Alice, and then sends the other particles to her after she has received all the first particles, and tells her the state of these Bell states.
In Step 2, Alice randomly forwards the particles of decoy Bell states to Bob and records the position of it, or simply discards them.
Bob places these particles in quantum registers following the incoming sequence.
In step 3, Alice asks Bob to perform Bell measurement on particles of $l$ $(l<h)$ decoy Bell states, and Bob sends result to Alice.
In this case, Alice can check the result from Bob and the state information from the trusted center Trent.
If Bob spoils, he can get through Alice's detection with the probability $\frac{1}{4}$ of every Bell
state.
He will be detected if Alice finds that at least one output is incorrect, so the probability of an incorrect output being accepted is $\frac{1}{4^l}$.

\section{Universal single-server and fully classical-client blind quantum computation protocol and new style for the blind quantum computation}

\subsection{Universal single-server and fully classical-client blind quantum computation protocol}

The client requires to get qubits from the trusted center and forwards them to quantum server in our protocol above.
The client can be fully classical if the first three steps are changed as follows.

\textbf{Step $1^{'}$} Trent prepares $2n$ Bell states $|\psi_{0,0}(B_k,A_k)\rangle (k=1,2,\cdots,2n)$, where $n=(2+\delta)m$ ($\delta>0$ is fixed number), and distributes the particles $B_k$ of each Bell state to Bob.
After Bob gets all the particles, Trent randomly selects particles from the second particles $A_k$ $(k=1,2,\ldots,2n)$ and discards them, and sends the rest particles of $A_k$ to the server Bob.

\textbf{Step $2^{'}$} The trusted center notifies Alice which particles of $A_k$ are discarded and which are sent to Bob, and keeps this information secret to Bob.

\textbf{Step $3^{'}$}
Suppose that the $2m$ particles are $A_{s_1},A_{s_2},\cdots,A_{s_{m}}$ and $A_{t_1},A_{t_2},\cdots,A_{t_{m}}$, where $1\leq s_i\leq n<t_i\leq 2n, i\in\{1,2,\cdots,m\}$.
Alice notifies Bob to perform Bell state measurements on particles $A_{s_i}$ and $A_{t_j}$, $(i,j)\in\{1,2,\cdots,m\}$ and gives their positions in the sequence of the particles which are from Alice.
Bob implements measurement on the particles and submits the outcome $(z_{s_i}^{'},x_{t_j}^{'})\in \{0,1\}^2$ to Alice.

In the above modified protocol, the trusted center Trent instead of Alice sends the second particles of Bell states to the server and just notify Alice which particles are forwarded to the server and which are discarded. During this procedure, Alice can achieve quantum computation without the ability to possess quantum memory and access quantum channel.

\subsection{New style for the blind quantum computation}
Since it is believed to be impossible for the blind quantum computation with only classical client and single (untrusted) server~\cite{Dunjko2016,TMTK,Morimae2019}, the trusted center is essential if the client is classical and only one server is involved.
The role of the trusted center in the future BQC protocol might be similar to Certificate Authority(CA) in ``E-commerce" today \cite{Adams1999,2002Building}.
CA is a trusted third party trusted both by the party who is the subject of the certificate and by the party who relies on the certificate, while the trusted center in BQC protocol can be treated as a third party trusted both by classical clients and quantum servers.
Therefore, the blind quantum computation may work in ``Cloud + CA" style in the future, our single-server and almost-classical-client or single-server and classical-client blind quantum computation protocols might becomes a key ingredient of real-life application because the blind quantum computation is unconditionally security. 

\section{Conclusions}
\label{conclusions}
In conclusion, we propose a universal single-server BQC protocol for the classical client. In our protocol, the client is almost classical. In other words, he doesn't require any quantum technology or resources, and only receives and sends qubits. 
Furthermore, our protocol can delegate quantum computation to one quantum server, while keeping the input, output and algorithm private.
The client will be fully classical if the trusted center helps the client send particles to the server and tells the client which particles are sent or discarded.

The client interfaces with only one server in our schemes instead of three servers in the triple-server protocol \cite{Qinli}. Although all the three servers can be treat as one big server since communication among servers is allowed, it will make the BQC procedure complex for client to interact with three servers and the limitation of quantum resource may not allow us to possess plenty of quantum servers. Our protocols are as secure as the triple-server protocol, and that only one server makes it easier to implement and more practical.
Since it is believed to be impossible for blind quantum computation with only classical client and single (untrusted) server,
blind quantum computation may work in the ``Cloud + CA"
style in the future. To consider the practical application, the ``Cloud + CA" style may be worked for blind quantum computation in the first generation of quantum computations. Our protocols might become a key ingredient for real-life application in the first generation of quantum computations.

\begin{acknowledgments}
We are grateful to Guang Ping He, Anne Broadbent, Simone Severini and Qin Li, especially Kejia Zhang for revising the original manuscript and helpful discussions and suggestions. We would also like to thank the anonymous referee for his/her
helpful comments and suggestions to improve the original manuscript. We carried on this partial work
while Wang was an academic visitor of
the Department of Computer Science and the Department of Physics \& Astronomy
at University College London.
This work is (was) supported by the National Natural Science Foundation of China under Grant Nos. 62072119, 61672007, and 61272013, and by Guangdong Basic and Applied Basic Research Foundation under Grant No. 2020A1515011180.
\end{acknowledgments}



\begin{thebibliography} {99}

\bibitem{PWSHOR}
P. W. Shor, Algorithms for quantum computation. In:Proceedings of the 35th Annual Symposium on Foundations of Computer Science. {\it Santa Fe: IEEE Computer Society Press}, pp. 124. (1994).

\bibitem{LKG} L.K. Grover, A fast quantum mechanical algorithm for database search. In: Proceedings of the Twenty-Eighth Annual ACM Symposium on Theory of Computing (ACM) pp. 212--219. (1996).
\bibitem{Nielsen}
M. A. Nielsen and I. L. Chuang, \emph{Quantum Computation and Quantum Information} (Cambridge, Cambridge University
Press, 2000).

\bibitem{Childs}
A. M. Childs, Secure assisted quantum computation. \emph{Quantum Inf. Comput.} \textbf{5}, 456-466 (2005).

\bibitem{BFK}
A. Broadbent, J. Fitzsimons, \& E. Kashefi, Universal blind quantum computation. \emph{Proc. of the 50th Annual IEEE Sympo. on Found. of Comput.
Sci.} 517-526 (2009).


\bibitem{BKBFZW}
S. Barz, E. Kashefi, A. Broadbent, J. F. Fitzsimons, A. Zeilinger, and P. Walther, Demonstration of Blind Quantum Computing. {\it Science} \textbf{335}, 303-308 (2012).


\bibitem{MFujii}
T. Morimae, \& K. Fujii, Blind topological measurement-based quantum computation. {\it Nat. Commun.} \texttt{3}, 1036 (2012).

\bibitem{MVFMBQC}
T. Morimae, Verification for measurement-only blind quantum computing. {\it Phys. Rev. A} \textbf{89}, 060302(R) (2014).



\bibitem{MDF}
A. Mantri, C. A. P. Delgado, \& J. F. Fitzsimons, Optimal blind quantum computation. {\it Phys. Rev. Lett.} \textbf{111}, 230502 (2013).


\bibitem{TMKF}
T. Morimae \& K. Fujii, Secure Entanglement Distillation for Double-Server Blind Quantum Computation. {\it Phys. Rev. Lett.} \textbf{111}, 020502 (2013).




\bibitem{Qinli}
Q. Li, W. H. Chan, C. Wu, \& Z. Wen, Triple-Sever blind quantum computation using entanglement swapping. {\it Phys. Rev. A} \textbf{89}, 040302(R) (2014).

\bibitem{Sheng} Y.-B. Sheng, L. Zhou. Deterministic entanglement distillation for secure double-server blind quantum computation. {\it Scientific Reports}, \textbf{5},7815 (2015).

\bibitem{Hayashi2015}
M. Hayashi and T. Morimae, Verufuabke measurement-only blind quantum computing with stabilizer testing. {\it Phys. Rev. Lett.} \textbf{115}, 220502 (2015).

\bibitem{Morimae2015}
T. Morimae, V. Dunjko, E. Kashefi. Ground state blind quantum computation on AKLT state \emph{Quantum Inf. Comput.} \textbf{15}, 200-234 (2015).

\bibitem{Greganti2016}
C. Greganti, M. Roehsner, S. Barz, T. Morimae, P. Walther. Demonstration of
measurement-only blind quantum computing, {\it New J. Phys.} \textbf{18}, 013020 (2016).

\bibitem{Kong2016} X. Kong, Q. Li, C. Wu, F. Yu, J. He, Z. Sun, Multiple-server Flexible Blind Quantum Computation
in Networks. {\it Int. J. Theor. Phys.}, \textbf{55},3001 (2016).

\bibitem{Fitzsumons2017}
J. F. Fitzsumons and E. Kashefi, Uncondifionally verifiable blind computation. {\it Phys. Rev. A} \textbf{96}, 012303 (2017).
\bibitem{Sheng2018}
Y.-B. Sheng and L. Zhou, Blind quantum computation with a noise channel. {\it Phys. Rev. A} \textbf{98}, 052343 (2018).

\bibitem{Li2018}
Q. Li, Z. Li, W. H. Chan, S. Zhang, C. Liu, Blind quantum computation with identity authentication. {\it Phys. Lett. A} \textbf{382}, 938 (2018).

\bibitem{fitzsimons2017private}
Fitzsimons, Joseph F, Private quantum computation: an introduction to blind quantum computing and related protocols. {\it npj Quantum Information} \texttt{3}, 1 (2017).

\bibitem{Dunjko2016}
V. Dunjko and E. Kashefi, Blind quantum computing with two almost identical states. arXiv:1604.01586

\bibitem{TMTK}
T. Morimae, T. Koshiba, Impossibility of perfectly-secure one-round delegated quantum computing for classical client \emph{Quantum Inf. Comput.} \textbf{19}, 0214-0221 (2019).

\bibitem{Morimae2019}
T. Morimae, H. Nishimura, Y. Takeuchi, S. Tani, Impossibility of blind quantum sampling for classical client \emph{Quantum Inf. Comput.} \textbf{19}, 0793-0806 (2019).



\bibitem{briegel2009measurement}
Briegel, Hans J and Browne, David E and D{\"u}r, Wolfgang and Raussendorf, Robert and Van den Nest, Maarten, Measurement-based quantum computation. {\it Nature Physics} \texttt{5}, 19 (2009).

\bibitem{jozsa2006introduction}
Jozsa, Richard, An introduction to measurement based quantum computation. {\it NATO Science Series, III: Computer and Systems Sciences. Quantum Information Processing-From Theory to Experiment}(Amsterdam: IOS Press) \texttt{199}, 137 (2006).

\bibitem{raussendorf2003measurement}
Raussendorf, Robert and Browne, Daniel E and Briegel, Hans J, Measurement-based quantum computation on cluster states. {\it Phys. Rev. A} \textbf{68}, 022312 (2003).

\bibitem{reichardt2013classical}
Reichardt, Ben W and Unger, Falk and Vazirani, Umesh, Classical command of quantum systems. {\it Nature} \texttt{496}, 7446 (2013).

\bibitem{huang2017experimental}
Huang et al., Experimental blind quantum computing for a classical client. {\it Phys. Rev. Lett.} \textbf{119} 050503 (2017).

\bibitem{Zukowski}
M. Zukowski, A. Zeilinger, M. A. Horne, and A. K. Ekert, "Event-ready-detectors`` Bell experiment via entanglement swapping. {\it Phys. Rev. Lett.} \textbf{71}, 4287 (1993).

\bibitem{Pan}
J.-W. Pan, D. Bouwmeester, H. Weinfurter, and A. Zeilinger, Experimental Entanglement Swapping: Entangling Photons That Never Interacted. {\it Phys. Rev. Lett.} \textbf{80}, 3891 (1998).

\bibitem{Bose}
S. Bose, V. Vedral, \& P. L. Knight, Multiparticle generalization of entanglement swapping. {\it Phys. Rev. A} \textbf{57},822 (1998).

\bibitem{Polking}
R. E. S. Polkinghorne, \& T. C. Ralph, Continuous Variable Entanglement Swapping. {\it Phys. Rev. Lett.} \textbf{83}, 2095 (1999).

\bibitem{Ma}
X. Ma, S. Zotter, J. Kofler, R. Ursin, T. Jennewein, C. Brukner, and A. Zeilinger, Experimental delayed-choice entanglement swapping. {\it Nat. Phys.} \textbf{8}, 479 (2012).




\bibitem{Bennett}
C. H. Bennett, G. Brassard, C. Crepeau, R. Jozsa, A. Peres, and W. K. Wootters, Teleporting an unknown quantum state via dual classical and Einstein-Podolsky-Rosen channels. {\it Phys. Rev. Lett.} \textbf{70} 1895 (1993).

\bibitem{Lu}
H. Lu and G. Guo, Teleportation of a two-particle entangled state via entanglement swapping. {\it Phys. Lett. A} \textbf{276}, 209 (2000).

\bibitem{Zhang}
Z. J. Zhang, and Z. X. Man, Multiparty quantum secret sharing of classical messages based on entanglement swapping. {\it Phys. Rev. A} \textbf{72}, 022303 (2005).

\bibitem{Xiu}
X. M. Xiu, H. K. Dong, Y. J. Gao, and F. Chi, Deterministic secure quantum communication using four-particle genuine entangled state and entanglement swapping. {\it Opt. Commum.} \textbf{282}, 2457 (2009).

\bibitem{Briegel}
H. J. Briegel, W. Dur, J. I. Cirac, and P. Zoller, Quantum Repeaters: The Role of Imperfect Local Operations in Quantum Communication. {\it Phys. Rev. Lett.} \textbf{81} 5932 (1998).

\bibitem{Duan}
L. M. Duan, M. D. Lukin, J. I. Cirac, and P. Zoller, Long-distance quantum communication with atomic ensembles and linear optics. {\it Nature} (London) \textbf{414}, 413(2001).


\bibitem{Adams1999} Carlisle Adams and Steve Lloyd, \emph{Understanding public-key infrastructure: Concepts, Standards, and Deployment Considerations} (Indianapolis: New Riders Publishing, 1999).

\bibitem{2002Building}
Atif, Y. Building Trust in E-Commerce. {\it IEEE Internet Computing} \textbf{6}, 18-24 (2002).

\end{thebibliography}

\end{document}